\documentclass[useAMS,usenatbib,usegraphicx]{mn2e}
\newcommand{\kms}{\, \rm{km}\,  \rm{s}^{-1}}

\def\ltsima{$\; \buildrel < \over \sim \;$}
\def\lsim{\lower.5ex\hbox{\ltsima}}

\def\mpch{\mbox{$h^{-1}$Mpc}}

\def\Mpc{{\rm Mpc}}

\def\deg{\ifmmode{^\circ}\else{$^\circ$}\fi}
\def\hGpc{\ifmmode{h^{-1}{\rm Gpc}}\else{$h^{-1}{\rm Gpc}$}\fi}
\def\hkpc{\ifmmode{h^{-1}{\rm kpc}}\else{$h^{-1}{\rm kpc}$}\fi}
\def\hMpc{\ifmmode{h^{-1}{\rm Mpc}}\else{$h^{-1}{\rm Mpc}$}\fi}
\def\hMsun{\ifmmode{h^{-1}M_\odot}\else{$h^{-1}M_\odot$}\fi}

\def\kms{{{\rm km}/{\rm s}}}

\def\muK{\ifmmode{\mu{\rm{K}}}\else{$\mu$K}\fi}
\def\mum{\ifmmode{\mu{\rm{m}}}\else{$\mu$m}\fi}

\newcommand{\LCDM}{$\Lambda$CDM}
\newcommand{\aj}{{AJ}}
\newcommand{\apj}{{ApJ}}

\newcommand{\apjs}{{ApJS}}
\newcommand{\mnras}{{MNRAS}}

\title{The Properties of Galaxies in Voids}

\author[Patiri et al.]{
\parbox[t]{\textwidth}{Santiago G. Patiri$^{1}$\thanks{E-mail:
spatiri@iac.es},
Francisco Prada$^{2}$, 
Jon Holtzman$^{3}$,
Anatoly Klypin$^{3}$ \&
Juan Betancort-Rijo$^{1,4}$
}
\\
\\
$^1$
Instituto de Astrofisica de Canarias,
C/ Via Lactea s/n, Tenerife, E38200, Spain
\\
$^2$
Ramon y Cajal Fellow, Instituto de Astrofisica de Andalucia (CSIC), E-18008, Granada, Spain
\\
$^3$
Astronomy Department,
New Mexico State University, Dept.\ 4500,
Las Cruces, NM 88003, USA
\\
$^4$
Facultad de Fisica, Universidad de La Laguna,
Astrofisico Francisco Sanchez, s/n, La Laguna
Tenerife, E38200, Spain
\\
}

\begin{document}

\maketitle

\begin{abstract}

We present a comparison of the properties of galaxies in the most
underdense regions of the Universe, where the galaxy number density is
less than 10\%  of the mean density, with galaxies from more typical regions.
We have compiled a sample of galaxies in 46 large nearby voids that
were identified using the Sloan Digital Sky Survey DR4, which provides
the largest coverage of the sky. We study the $u-r$ color distribution,
morphology, specific star formation rate, and radial number density profiles
for a total of 495 galaxies fainter than $M_{r}=-20.4 +5\log h$
located inside the voids and compare these properties with a control
sample of field galaxies. We show that there is an excess
of blue galaxies inside the voids. However, inspecting the properties
of blue and red galaxies separately, we find that galaxy properties
such as color distribution, bulge-to-total ratios, and concentrations
are remarkably similar between the void and overall sample.  The void
galaxies also show the same specific star formation rate at fixed color as
the control galaxies.  We compare our results with the predictions of
cosmological simulations of galaxy formation using the Millennium Run
semi-analytic galaxy catalog. We show that the properties of the
simulated galaxies in large voids are in reasonably good agreement
with those found in similar environments in the real Universe.  To
summarize, in spite of the fact that galaxies in voids live in the
least dense large-scale environment, this environment makes very little
impact on properties of galaxies.
\end{abstract}


\section{Introduction}

It has been known for many years that the properties of the galaxies
depend on their local environments. Dressler (1980) pointed out that
the fraction of late-type spirals is larger in regions of lower local
density. A similar trend was observed for star formation rates (SFR):
star-forming galaxies lie in more underdense regions.  On the other
hand, the most massive ellipticals and galaxies with low star formation
rates reside preferentially in the most dense environments, especially
in rich clusters of galaxies (e.g. Postman \& Geller 1984; Dressler et
al. 2004). Recently, Hogg et al.(2004)  show that in all different 
local environments the bulge-dominated galaxies, defined as those with large
S\'ersic indices, show a color-magnitude diagram dominated by red galaxies 
and that the most luminous galaxies reside preferentially in high-density
regions while blue galaxies live mainly in low-density regions.

The $u-r$ color distribution drawn from the Sloan Digital Sky Survey
(SDSS) is well modeled by two Gaussian distributions which can be
used to split the samples into a blue and a red galaxy population
(see Strateva et al. 2001; Baldry et al. 2004). Although the mean and
variance of these two distributions are strong functions of luminosity,
or stellar mass (see also Bernardi et al. 2003a; Blanton et al. 2003b;
Hogg et al. 2004), there is only a weak dependence of these quantities
on the environment (Balogh et al. 2004; Hogg et al. 2004). Nevertheless,
the relative fraction of red and blue galaxies strongly depends on local
density. At fixed luminosity, the fraction of red galaxies change
from 10\%-30\% in low density environments to about 70\% in highest density 
environments.

Extensive studies on the dependence of galaxy properties with environment
has been done for galaxies lying in high-density regions such as groups
and clusters and its comparison with the field (see Dressler et al. 2004,
and references therein). More recently, thanks to the advent of large
redshift surveys, it has become possible to study the properties of
galaxies lying in the most underdense regions of the Universe.

Grogin \& Geller (2000) selected a sample of galaxies from the CfA
redshift survey that are located in regions with galaxy number density contrast 
($\delta_{gal}=(\bar{n}_{v}/\bar{n})-1$, where $\bar{n}_{v}$ is the number
density of galaxies in the scale of interest and $\bar{n}$ is the mean number
density of galaxies in the sample) of $\leq -0.5$. 
They studied a total of 149 galaxies in these environments, finding these 
galaxies to be bluer than those in the field and also showed that a significant 
fraction of them were late type spirals.  Recently, Rojas et al. (2004, 2005) 
studied a sample of $\sim 1200$ galaxies, extracted from the SDSS SAMPLE10 
(about 1.5 times the SDSS DR1), located in regions with $\delta_{gal} \leq -0.5$.  
They also find that void galaxies are bluer on average than the general
galaxy population and that they have a higher specific star formation
rate. They claim that this difference is not just caused by the
morphology-density relation; they find that late-type galaxies (as defined
by a global Sersic index fit) in voids differ from those outside of voids.
Croton et al. (2005), using the 2dFGRS, also showed that the galaxy
population in underdense regions differs from that in the field. They
found that a late-type population dominates the low density regions, while
cluster regions have a relative excess of very bright early-type galaxies.

The study of the properties of galaxies in large underdense regions can
provide strong constrains to the galaxy formation processes. Benson et
al.(2003) used semi-analytic models in order to study the predictions
concerning galaxies within voids. Although the resolution they used in
their simulations were not sufficiently high to take full description of
the properties of galaxies in underdense regions, they found that the
galaxy population in these regions is bluer than the field. Also, they
found that the specific star formation rate is higher in the central
regions of the voids than the boundaries. Today, with the advent of
higher resolution simulations and the improvements in the prescriptions
for galaxy formation it is possible to make more robust predictions.

In this paper, we use the latest public data release of the SDSS (Data
Release 4) in order to study statistically the properties of galaxies
lying in the most underdense regions of the universe, those with 
$\delta_{gal}$ about -0.9. We specifically take
advantage of a large contiguous region spanned by the DR4 in order 
to find for the first time large voids that have previously been impossible 
to identify. In addition, we compare our results with the latest predictions 
from the semi-analytic models, using the recently released Millennium Run 
galaxy catalog (Springel et al. 2005; Croton et al. 2005). In order to
make this comparison, we analyze the galaxies found in voids in 
this mock catalog.  


The work is organized as follows. In Section 2 we describe the
observational data set and the method for searching for large voids. In
Section 3 we present our results on the properties of galaxies in 
voids. In Section 4 we compare our results with the predictions of 
the recently released Millennium Run galaxy
catalog (Springel et al. 2005; Croton et al. 2005). 
Finally, in Section 5, we present our conclusions.

Throughout the paper we assume a \LCDM~ cosmology with parameters $\Omega_m=0.3$, 
$\Omega_{\Lambda}=0.7$ and $H_{0}=100 h \kms/\Mpc $.

\section{Searching for Voids in the SDSS DR4}
\label{sec:samples}

The SDSS is the largest photometric and spectroscopic survey to
date; the completed survey will cover approximately 8000 square deg. 
CCD imaging is done in five colors and spectroscopic follow-up of galaxies 
is performed down to r=17.77 mag; see York et al.(2000) and Stoughton
et al.(2002) for a complete overview of the survey. The details of
selection function for galaxy spectroscopy and spectroscopic tiling can
be found in Strauss et al.(2002) and Blanton et al.(2003a). A thorough
analysis of systematic uncertainties in the galaxy samples is described
in Scranton et al. (2002).

Although the SDSS spectroscopic sample is magnitude limited, there is some
incompleteness due to missing nearby bright galaxies ($r < 15.5$) to avoid fiber 
saturation, and missing galaxies from the need to avoid fiber collisions (7\%; Blanton et al. 2003a), 
redshift failures ($<$1\%) and some galaxies ($\sim$1\%) blotted out by 
bright Galactic stars. This incompleteness may cause a slight overestimation
in the final underdensities.

In the present work we  use the latest public version of the
survey, the SDSS DR4. This release has a total of 410,939 galaxy spectra
distributed in several regions or strips covering a region of about
4,800 sq.degrees on the sky. The galaxy sample is distributed in three
regions. The first one is located in the North Galactic Cap (NGC) and
the other two are in the South Galactic Cap (SGC), near the celestial
equator. Although the geometry of these regions is rather complex,
we were able to select a geometrically contiguous region from the NGC,
which we refer to as the North Strip (NS). The North Strip covers an area
of $\sim$2,000 sq. degrees on the sky going from $117\deg < {\rm RA} < 200\deg$
and $+20\deg < {\rm DEC} < +70\deg$. The large declination range of this region
allows us to search for large nearby voids with high underdensities.

We define voids as the maximal non-overlapping spheres empty
of galaxies with luminosity above a given value. For example, we could
define voids as maximal spheres empty of Milky Way-size galaxies. Although
the voids are empty of these galaxies, they can have fainter galaxies
inside (see Patiri et al. 2006b for a full discussion on different void
definitions and uses). Note that as we are interested in studying the
galaxies located in the most underdense regions of the universe we 
search for very large voids, i.e. those with a radius much larger
than the mean distance between galaxies.

To select galaxies which  define our voids, we use SDSS {\it r}-band 
magnitudes. 
Absolute magnitudes are computed from the extinction-corrected apparent 
magnitudes assuming a $\Lambda$CDM cosmology and taking into account the 
{\em k}-correction using the prescription developed by Blanton et al.(2003b).

The SDSS DR4 is a flux-limited survey. So, there is a decrease of observed
galaxies with redshift. In order to search for large nearby voids, we
 construct from the NS sample a volume-limited galaxy sample with
$z_{max}=0.085$ using redshifts with confidence level greater than 0.95,
corresponding to a volume out to a distance of $250 \mpch$. Within this
volume, the spectroscopic survey is complete to a limiting absolute
magnitude of $M_{{\it r}}^{lim}= -19.4 +5{\rm log} h$. The total
volume of our sample is $2.31 \times 10^{6} \mpch^{3}$ with a total
of 27046 galaxies.  

We search for large voids in our volume-limited galaxy sample using
the {\em HB void finder} (Patiri et al. 2006b). The {\em HB Void Finder}
is conceptually simple, as it searches for the maximal non-overlapping
spheres with radius {\em larger} than a given value that are empty of
galaxies brighter than some specified brightness. This code is designed
to be accurate and computationally efficient in finding the largest voids
in a galaxy sample. To search for voids, the code first generates a sphere
of fixed radius randomly located within the galaxy sample. We check
whether this sphere is empty of bright galaxies. If so, we increase
the size of the sphere until its surface reaches the four nearest galaxies. 
If this sphere does not overlap another one, we define it as a void. If 
it does overlap, the larger sphere is taken to be the void. 
To assure that the full sphere is entirely inside the galaxy sample, we put 
artificial guard ``luminous galaxies'' on the borders of the sample.
For a full description of the algorithm see Patiri et al.(2006b).

\begin{figure*}
\includegraphics[width=165mm]{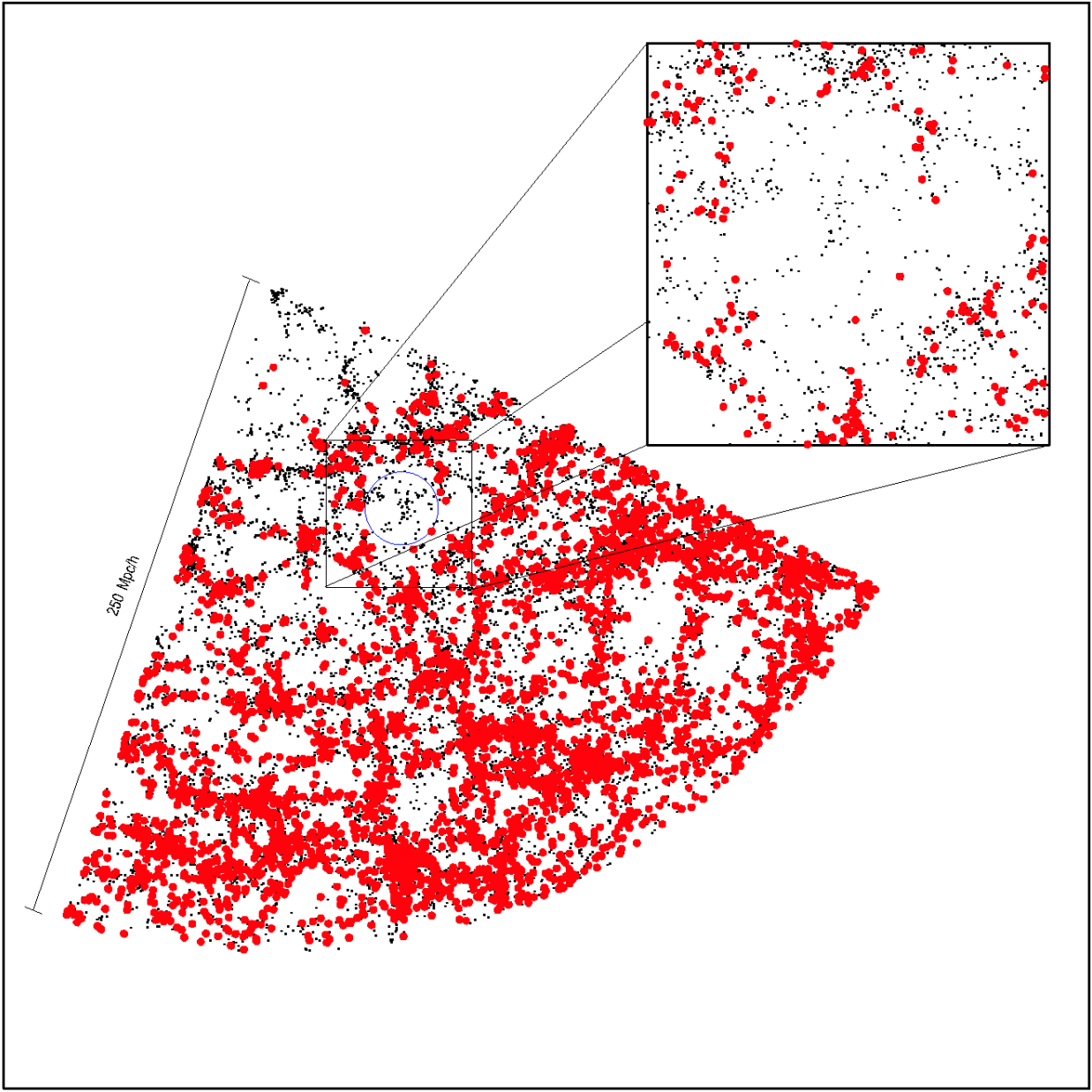}
\caption{A slab of $9 \mpch$ of our SDSS NS flux-limited galaxy sample. We
show one of our large voids (circle). This void has a radius of $13\mpch$
and it is defined by galaxies brighter than $M_{{\it r}}=-20.4 +5{\rm
log} h$ (big dots). We also plot fainter galaxies (small dots). The
inset shows a zoom of the void region of size $60 \mpch$. Note that
the faint galaxies populating the voids are not randomly distributed:
the are clustered in small groups and filaments in a similar way to that
of the brighter galaxies.}
\label{fig:fig0}
\end{figure*}

In the present paper we  search for voids defined by absence of
galaxies brighter than $M_{{\it r}}=-20.4 +5{\rm log} h$.
The mean number density of these galaxies is $4.2 \times
10^{-3} \mpch^{-3}$, with a mean distance between the galaxies of $\sim
6.2 \mpch$. Since we are interested in large voids, we  start the
void finder with a minimum radius of $10.0 \mpch$. 
Note, that a randomly placed sphere with a radius of $14.0 \mpch$ would contain
typically 50 galaxies of those defining the voids (i.e. galaxies
brighter than $M_{{\it r}}=-20.4 +5{\rm log} h$).

Using these definitions, we find in our volume-limited sample 
a total of 46 voids with radius larger than $10.0 \mpch$\footnote{The full 
list of coordinates and sizes of these voids are available upon request}. 
The largest void has a radius of $17.3\mpch$. In Figure 
\ref{fig:fig0} we show one of our large voids in the SDSS NS flux-limited sample 
and its surrounding large scale structure. The big dots denote the galaxies 
defining the voids (brighter than $M_{{\it r}}=-20.4 +5{\rm log} h$); 
small dots are fainter galaxies. In the inset we magnify the void region. Note that
although the void is empty of bright galaxies, it is populated by fainter
galaxies. These faint galaxies are not randomly distributed: they are
arranged in small groups and filaments. This distribution resembles the
large scale structure of the Universe. Very similar results were found in
the high resolution N-body simulations of voids defined by dark matter 
halos (Gottlober et al. 2003).

As we mentioned above, the extensive coverage of the most recent SDSS 
data release enables the search for such large empty voids for the first 
time. In previous studies of ``void galaxies'' in SDSS (Rojas et al. 2004, 2005), 
the data set was  much more limited. As a result, 
 ``void galaxies'' were selected by a nearest-neighbor analysis by looking for galaxies 
(within a specified magnitude range) whose third nearest neighbor was at distance larger 
than $7h^{-1}$~Mpc. This method includes galaxies in voids much smaller  than we 
consider here, and also occasionally allows galaxies in low-density filaments 
to be counted as void galaxies. Although this procedure yields a very large 
sample of galaxies, the environment where galaxies are located is not so underdense.
Our current sample represents the galaxies in the largest and emptiest
voids, and, thus, perhaps is expected to provide the strongest constraints
on how void galaxies compare to galaxies in the general population.

\section{The physical properties of galaxies in large voids}
\label{sec:res}

In our sample of 46 nearby voids larger than $10.0 \mpch$ we find a
total of 495 galaxies fainter than $M_{{\it r}}=-20.4 +5{\rm log} h$,
and brighter than our limiting magnitude of $M_{{\it r}}=-19.4 +5{\rm
log} h$.  To determine whether the properties of these galaxies differ
systematically from those outside of voids, using our volume-limited
sample we construct a control sample of field galaxies by taking
galaxies which happen to be inside randomly selected spheres of radius
$13\mpch$. We then take all galaxies in the same luminosity range as
our void galaxies. 
Using the void and the control samples, we can compare color distributions,
morphology, and specific star formation rates. We also investigate the
number density profiles of galaxies within voids. Finally, we compare the
observed properties of our void galaxies with simulated galaxies.

\subsection{Galaxy number density profiles}

We define the number density profile of galaxies in voids as:
\begin{equation}
1+\delta_{gal}(r)=\frac{n_{void}(r)}{n},
\end{equation}
where $n_{void}(r)$ is the mean number density of galaxies averaged over all our
voids at a distance $r$ from the center of the void and $n$ is the mean number density of 
galaxies in our volume-limited sample for the same magnitude bin. 

\begin{figure}
\includegraphics[width=\columnwidth]{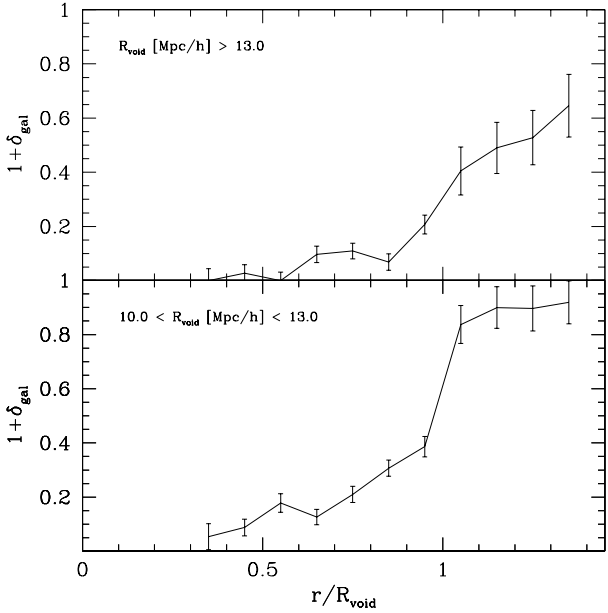}
\caption{Top panel: number density profile ($1+\delta_{gal}$)
for galaxies in voids with radius larger than $13 \mpch$ 
found in the SDSS NS sample as a function of the void radius. 
Bottom panel: number density profile 
for galaxies in voids with radius between $10 \mpch$ and $13 \mpch$.}
\label{fig:fig2}
\end{figure}

In the bottom panel of Figure \ref{fig:fig2} we show the
number density profile of faint galaxies in voids with radii between 
$10 \mpch$ and $13 \mpch$ as a function of distance from the void center 
(normalized to the void radius, $R_{void}$).
The solid line in Figure \ref{fig:fig2} shows the $1+\delta_{gal}$ profile. 
We can see that the number density profile of galaxies inside voids is quite
flat. The mean (enclosed) number density of 
galaxies at $R_{void}$ is 10\% that of the field. Beyond $R_{void}$
the local number density profile shows a remarkably steep increase,
reaching a local density of 90\% of the mean density at 1.4 $R_{void}$.

In the top panel of Figure \ref{fig:fig2} we show the number density
profile of galaxies for voids with radius larger than $13 \mpch$. We
can see a similar behavior shown in the smaller voids.  However,
larger voids have a flatter profile and the increase with radius is
not so sharp. At distance $1.4R_{void}$, the density is about
60\% of the mean density. Note that we do not see a spike in the
number density around $R_{void}$ as predicted by numerical and analytical 
estimates of voids defined by dark matter halos (see e.g., Colberg et al. 2005; 
Patiri et al. 2006a). A possible reason for the absence of the spike could be 
due to redshift distortions, which smooth out the spike. Spatial bias may also 
be involved.

Error bars in the number density profiles
have been calculated using the Millennium
Run galaxy catalog, which is discussed in more detail in
Section \ref{sec:sim}. Using the simulated catalog we can
estimate the deviations of the profiles that includes the sampling error,
as well as cosmic variance.
In order to compute the error bars, we use a list of 
2000 voids found in the entire catalog (see Section 4.1). From this list, 
we  randomly select a set of 50 samples each 
containing 46 voids as in our SDSS sample. For each sample
we obtain the number density profile averaging over the 46 voids. 
We find comparable number of galaxies per void. So, we end up with a sample
of 50 profiles and we estimate the {\emph rms} deviation in each radial bin 
over the ensemble of the 50 profiles.

\begin{figure}
\includegraphics[width=\columnwidth]{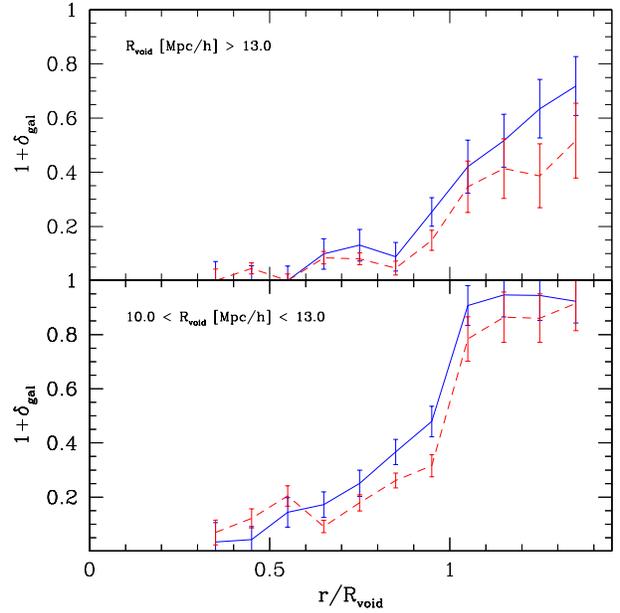}
\caption{Top panel: number density profile ($1+\delta_{gal}$)
for red galaxies (dashed line) and blue galaxies (full line) in voids 
with radius larger than $13 \mpch$. 
Bottom panel: the same profiles for voids with radii between $10 \mpch$ and $13 \mpch$.} 
\label{fig:fig2b}
\end{figure}

In Figure \ref{fig:fig2b}, we show the profiles for blue
(solid line) and red galaxies (dashed line) separately, for the smaller
voids (bottom) and larger voids (top). There are galaxies of both
populations throughout the voids; both red and blue galaxies increase
rapidly in density at the edges of the voids.

\subsection{Color distributions}

\begin{figure}
\includegraphics[width=\columnwidth]{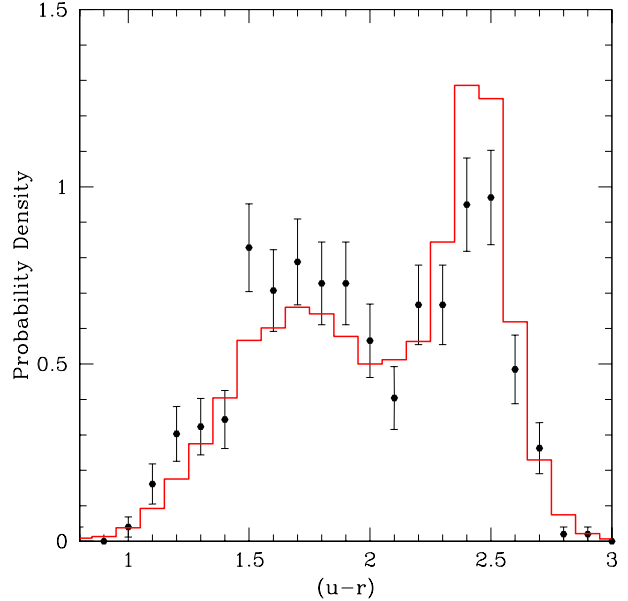}
\caption{$u-r$ color distribution for void galaxies found in 
the SDSS NS sample in the $-19.4 > M_{{\it r}}-5log h > -20.4$ magnitude range (black dots). 
The full line gives the $u-r$ color distribution for galaxies in
the control sample in the same magnitude bin. }
\label{fig:fig1}
\end{figure}

In Figure \ref{fig:fig1} we plot the $u-r$ color distribution of the
galaxies in our void sample (black points). We also show the 
color distribution of galaxies in the control sample (full line). 
The error bars ($\Delta F$) are derived using the binomial distribution:
\begin{equation}
\Delta F=\frac{\Delta n}{N} ;\quad \Delta n=\sqrt{n(1-F)},
\end{equation}
where $F$ is the galaxy number fraction for each $u-r$ color bin. $n$
is the number of galaxies in each bin, and $N$ is the total number of
galaxies found in our voids.

We clearly see that both $u-r$ galaxy color distributions are bimodal. There 
is an excess of blue galaxies in the voids relative to  
the control sample in this luminosity bin ($-19.4 >
M_{{\it r}}-5\log h > -20.4$). This behavior has been shown previously
for galaxies in local underdense regions with $\delta_{gal} < -0.6$ 
(e.g., Balogh et al. 2004; Rojas et al. 2004). 

\begin{figure}
\includegraphics[width=\columnwidth]{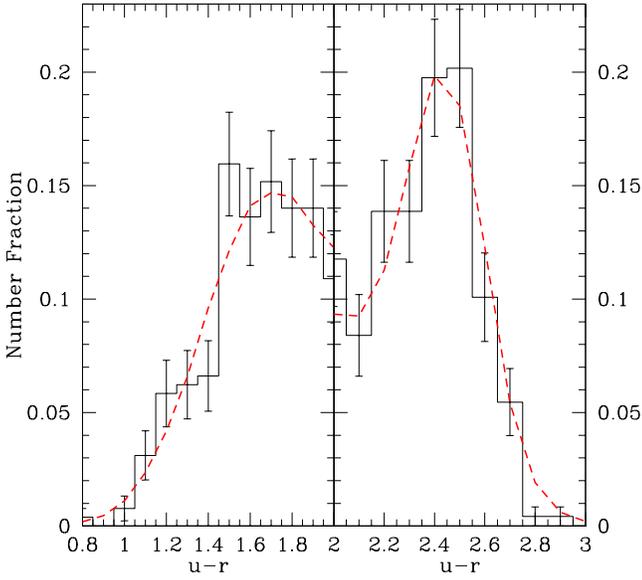}
\caption{Left panel: $u-r$ color distribution for galaxies lying in voids (full line) and
field galaxies (short dashed line) normalized to the blue population. 
Right panel: the same  but for the red population.
Note that the mean and variance of both populations are very similar 
for galaxies in voids and in the field.}
\label{fig:fig1b}
\end{figure}

However, if we look at the blue ($u-r < 2.0$) 
and red ($u-r > 2.0$) galaxy populations separately, we find 
a mean $<u-r>=1.61$ for the blue population in our voids,
while for the blue population in the control sample the mean color is 
$<u-r>=1.62$.  For red galaxies we get $<u-r>=2.37$ and
$<u-r>=2.4$ for void and field galaxies, respectively. The mean 
colors of red and blue galaxies separately are essentially identical 
inside voids and in the field. Figure \ref{fig:fig1b} shows
the normalized color distributions of blue and red galaxies
separately; it is clear that the void and field distributions
for galaxies of the same color are practically indistinguishable.

\subsection{Morphology}

While color is to some extent likely to be related to morphology,
our sample is sufficiently nearby that we are able to quantify
the morphology of galaxies directly from the SDSS imaging, and
to compare the morphology of void and control galaxies.

\begin{figure}
\includegraphics[width=\columnwidth]{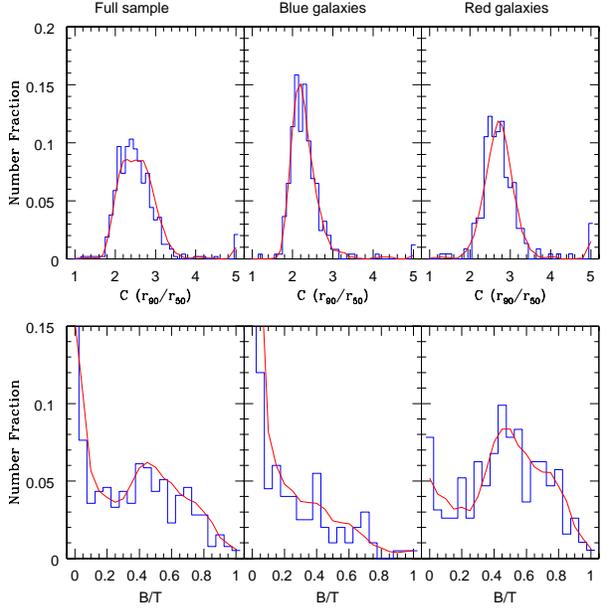}
\caption{Upper panels: Distribution of concentrations of the void (histograms) and field (curves) 
galaxies. In the three  panels we show the distributions for all the galaxies (left), 
for the blue population (center) and red population (right). 
Bottom panels: Distribution of Bulge-to-Total ratios for galaxies found in the SDSS. 
The histograms denote galaxies in the voids and the curves  are for  the galaxies in the field. 
Left panel shows all the galaxies, center panel only blue galaxies,
and right panel only red galaxies.}
\label{fig:cbt}
\end{figure}

Figure \ref{fig:cbt} (upper panels) shows the distribution of concentrations
of the void (histograms) and control (curves) samples; concentration is 
defined as $c\equiv r_{90}/r_{50}$, where $r_{90}$ and $r_{50}$
are the radii enclosing 90 and 50\% of the Petrosian flux respectively, 
as measured by the SDSS PHOTO pipeline. The comparison between
void and control samples is done for the entire samples (left
panel), as well as separately for the blue (center) and red (right)
galaxies. It is apparent that there are no visible differences
in the distribution of concentrations, especially within the
subsamples selected by color. Since blue galaxies have lower
concentrations than red galaxies, and since we have shown that
there are a larger fraction of blue galaxies in voids, the
concentrations for the full samples are shifted to
slightly smaller values in voids.

To get additional sensitivity to morphology, we perform bulge to
disk decomposition on all of the galaxies in the void and control
samples. This is done using a nonlinear least squares fit to
the data with a galaxy model that includes an inclined, circular,
exponential disk and a flattened Sersic bulge; additional
details are given in MacArthur et al. (2003). To do the fit,
a one-dimensional profile is constructed from the data by
fitting elliptical isophotes to the image. From the 1D profile,
an exponential disk model was fit from $r_{50} < r < r_{pet}$,
and a deVaucouleurs (Sersic $n=4$) profile was fit to $r < r_{50}$.
The 1D fits are done assuming a Gaussian PSF, where the FWHM was measured
from the SDSS PSF. These initial estimates are used as a starting guess
to fit a composite disk plus bulge model to the entire 1D profile out to
$r_{pet}$, fixing the bulge to be a deVaucouleurs bulge. The resulting
decomposition is then used as a starting guess for a 2D fit on the
SDSS data. The 2D fits are done using the 2D PSF determined by the
SDSS PHOTO pipeline. Separate fits are done fixing the Sersic index
at $n=1,2,3,4$ separately. Finally, a fit is done allowing for $n$ to
float, using the best fit model of the fixed Sersic fits as a starting
guess. We note that getting a unique fit for Sersic index is difficult
for most of the galaxies due to their limited angular extent; the
derived Sersic index has a strong covariance with the bulge effective
radius. We compute the bulge/total light ratio from the fits, where
$B/T$ is computed within the Petrosian radius; with this definition,
the $B/T$ ratio is rather insensitive to different combinations of
Sersic index and bulge effective radius.

In the bottom panels of figure \ref{fig:cbt}, we show the distribution
of $B/T$ ratios for the floating $n$ fit. Because of the insensitivity
of $B/T$ to the Sersic index, the distributions for the fixed and
floating Sersic index are very similar.  As with the concentrations,
the left panel shows the distributions for the entire void and control
samples, the center panel the distributions for the blue galaxies, and
the right panel the distributions for the red galaxies.  In this figure,
we can see that there are a significant number of galaxies with $B/T=0$.
These result for any galaxy for which an inward extrapolation of an
exponential fit to the outer regions matches or over-predicts the flux
in the central region. While pure disk galaxies with no bulge fall into
this category, so do spheroidal galaxies with an exponential profile (not
necessarily unexpected for the lower luminosity galaxies in our sample),
as do disk galaxies with profiles that are not fully exponential, e.g.
galaxies with an inner flattening or truncation. As a result, the $B/T=0$
bin is rather heterogeneous. Aside from this bin, the distributions of
$B/T$ ratios are quite similar between void and field galaxies. Although
the $B/T$ ratios are clearly different for the blue and red galaxies,
the relative change in the ratio of blue/red galaxies from field to void
is subtle enough that it is difficult to see the effect in the overall
distribution of $B/T$.

\begin{figure*}
  \includegraphics[width=165mm]{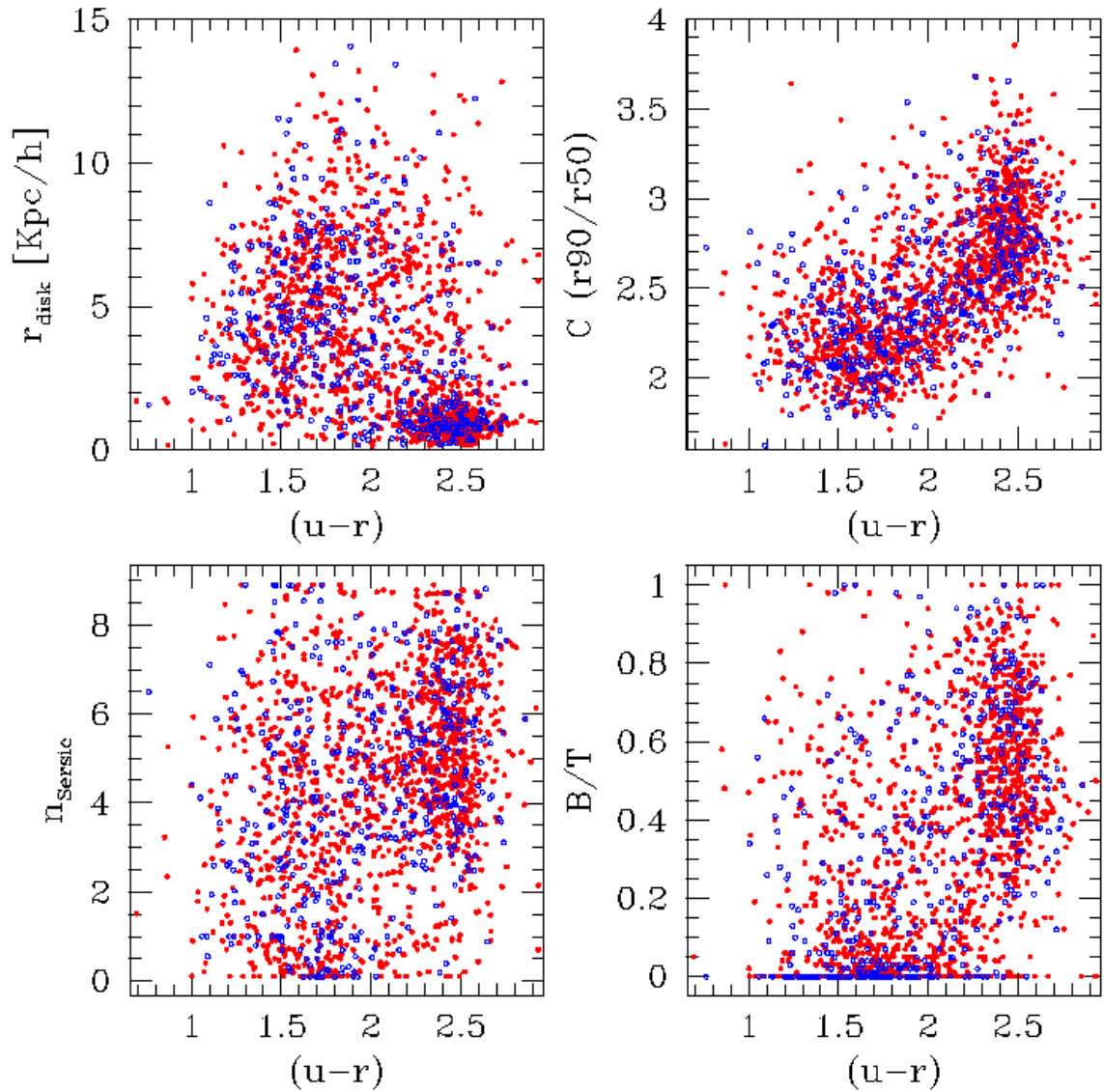}
\caption{Morphological properties of galaxies vs. $u-r$ color for
galaxies found in our SDSS sample. 
The open circles denote galaxies in voids and the full circles are the control galaxies.
Upper Left: Disk scale length. Upper Right: Concentration. Bottom Left: The Sersic 
parameter ($n$) obtained for the bulge. Bottom right: Bulge-to-Total luminosity ratios.}
\label{fig:panel1}
\end{figure*}

Figure~ \ref{fig:panel1} gives distribution of several morphological
properties of galaxies as function of color of the galaxies. There are no
apparent differences between void galaxies and the control sample in
any of the presented properties.

\subsection{Specific star formation rates}

The specific star
formation rate is defined as the star formation rate per unit of stellar
mass ($M_{*}$). The SFRs and stellar mass data for the galaxies in
the SDSS DR4 have been recently made publicly available\footnote{The
SFR and $M_{*}$ for the SDSS DR4 galaxies can be downloaded from
http://www.mpa-garching.mpg.de/SDSS/DR4/} (Kauffmann et al. 2003;
Brinchmann et al. 2004). The
SFRs were obtained by comparing the width of 
the emission lines with the continuum properties of the galaxies:
a method was developed
for aperture correction using the imaging available from the SDSS and
have shown that the method takes out essentially all aperture bias in the SFR
estimates, allowing an accurate estimate of the total SFRs in galaxies. 
(see Brinchmann et al. 2004 for details).
Stellar masses were estimated using the spectroscopic information and
broad-band photometry available in the SDSS (Kauffmann et al. 2003). 
The procedure also provides an estimate of the extinction.

\begin{figure}
\includegraphics[width=\columnwidth]{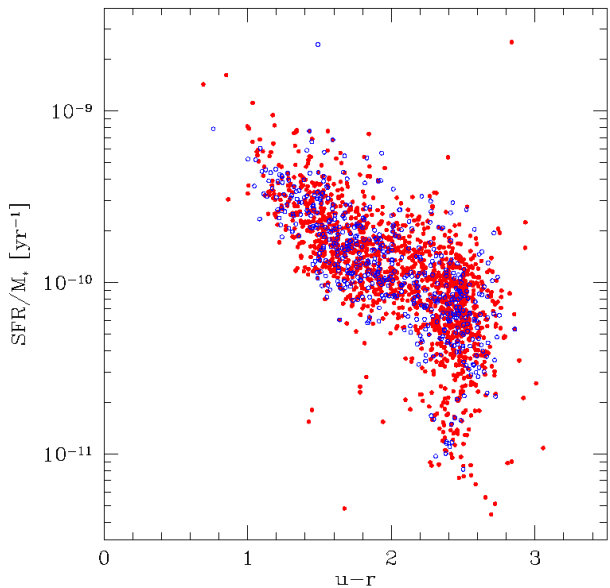}
\caption{The specific star formation rate of galaxies
lying in our large voids in the SDSS NS sample as a function of 
$u-r$ color (open circles). The full circles show the same for 
the galaxies in the field. There is not a
significant difference for galaxies of fixed color between
voids and field.}
\label{fig:fig4}
\end{figure}

Figure \ref{fig:fig4} shows the specific
star formation rate of our void galaxies as a 
function of $u-r$ color (open circles); data for the control sample
are also shown (full circles). There is a strong correlation between
galaxy color and specific SFR. Although there are relatively more
blue galaxies (with higher specific SFR) in voids, 
the void galaxies appear indistinguishable from the galaxies
in the control sample at fixed galaxy color. This suggests
a common star formation history inside and outside of
voids for galaxies with the same current star formation rate.

\section{Comparison with the Millennium Run}
\label{sec:sim}

\subsection{The galaxy catalog}

The Millennium Run is a large cosmological numerical simulation of
$2160^{3}$ dark matter particles in a box of $500 \mpch$ on a side;
for a full description see Springel et al. (2005b). The formation of
galaxies is tracked from the dark matter simulation using a semi-analytic
model to follow gas, stars and supermassive black hole processes within
the merger history trees of dark matter halos and their substructures.
The modeling assumptions and parameters are adjusted by trial and error in
order to fit the observed properties of low redshift galaxies, primarily
their joint luminosity-color distribution and their distributions of
morphology, gas content and central black hole mass.

The public available Millennium Run galaxy catalog\footnote{It can
be downloaded from http://www.mpa-garching.mpg.de/galform/agnpaper/}
(Springel et al. 2005; Croton et al. 2005) contains a total of about
9 million galaxies in the full simulation box. For each galaxy, the catalog
provides the position and velocity, the total and bulge galaxy magnitudes
in 5 bands (SDSS ugriz), the total and bulge stellar mass, cold,
hot and ejected gas mass, the black hole mass, and the star formation rate.

We use these data to study the properties of galaxies in the
most underdense regions of the simulated catalog. This offers an
unique opportunity to compare with the results that we  find in the
SDSS. We make the same analysis for the Millennium galaxy catalog as
 for our volume-limited galaxy sample drawn from the SDSS. To
this end, the first step is to search for voids. We divide the entire
Millennium box in several smaller boxes of comparable volume to that of
our SDSS NS galaxy sample; specifically, we  divide the $500\mpch$
box into 27 small boxes with $166.67 \mpch$ on a side. In order to make
a direct comparison with observations, we transformed the real space
catalogs produced for the small boxes to redshift space using the real
coordinates and the peculiar velocities.  Then, we search for voids in each
small box with the same parameters that we use for the SDSS analysis,
i.e. we searched for voids defined to be empty of galaxies brighter than
$M_{{\it r}}=-20.4 +5{\rm log} h$. The mean number density of these
galaxies is $4.3 \times 10^{-3} \mpch^{-3}$, very similar to that
found in our SDSS NS sample. On average over the small Millennium Run
boxes, we have found 78 voids larger than $10.0 \mpch$. This number is 
compatible with the number of voids found in the SDSS taking in account
that the small boxes in Millennium have twice the volume than our 
observational sample.

\subsection{Properties of the Millennium Run galaxies}

We study the properties of the fainter galaxies lying in these
voids down to $M_{{\it R}}=-19.4 +5{\rm log} h$ as we did in the
SDSS. We find on average $\sim 1000$ galaxies in the voids for each
box, which again is compatible with number of galaxies found in the
SDSS. 

\begin{figure}
\includegraphics[width=\columnwidth]{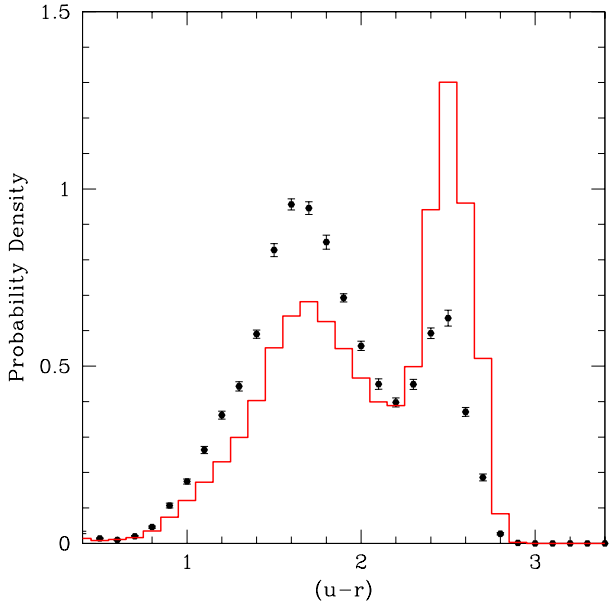}
\caption{$u-r$ color distribution for galaxies with 
$-19.4 > M_{{\it r}}-5{\rm log} h > -20.4 $ lying in large voids
found in the Millennium Run (black dots). The color distribution for
the galaxies in the field is also shown for comparison (full line).}
\label{fig:fig6}
\end{figure}

\begin{figure}
\includegraphics[width=\columnwidth]{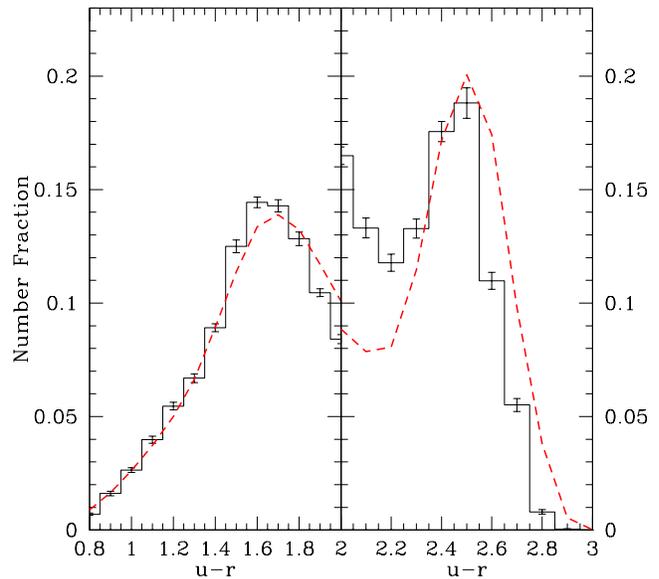}
\caption{Left Panel: $u-r$ color distribution for simulated blue galaxies 
lying in voids (full line) and field galaxies (short dashed line).
Right Panel: the color distribution of the red population. Here,
we can see that red population in voids is shifted compared with the field.}
\label{fig:fig6b}
\end{figure}

In Figure \ref{fig:fig6} we show the $u-r$ color
distribution of galaxies lying in these high underdense regions found in
all boxes (black dots). The error bars are the standard deviation derived
from the ensemble of the 27 small boxes. We also show the comparison for
galaxies in the field for the same magnitude interval (full line).
In the voids, there is clearly a larger ratio of blue to red
galaxies, as we see in the observations. However, the difference
between voids and field is clearly more pronounced in the simulations
than in the real data. The simulations have significantly more
blue galaxies than red galaxies in the voids, while the observations
show roughly comparable numbers.
If we divide the galaxies into blue and red populations, we find 
that the mean and variance of the blue population is the same in 
voids and in the field, as we see in the observations (see fig.\ref{fig:fig6b}).
Red galaxies in the simulations appear to differ slightly 
between voids and field, which is somewhat different from
what we see in the observations.

\begin{figure}
\includegraphics[width=\columnwidth]{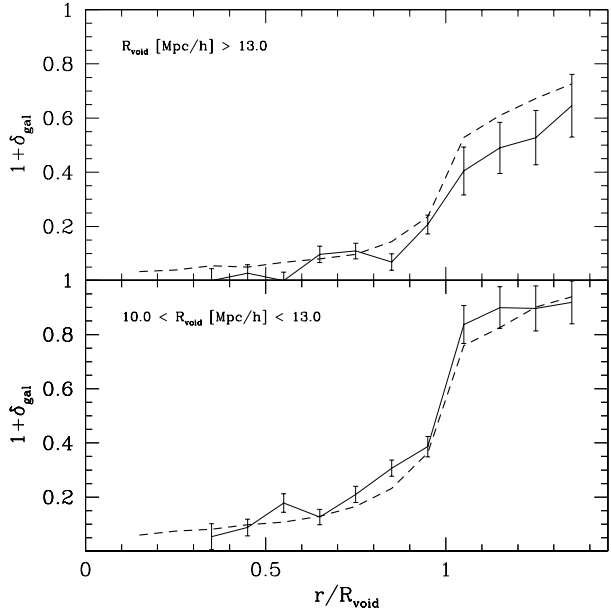}
\caption{Comparison between the number density profiles for voids
found in the Millennium Run (dashed line) and SDSS (full line). 
Top panel: profiles for voids larger than $13 \mpch$. 
Bottom panel: for voids with radius between $10 \mpch$ and $13 \mpch$}
\label{fig:fig7c}
\end{figure}

\begin{figure}
\includegraphics[width=\columnwidth]{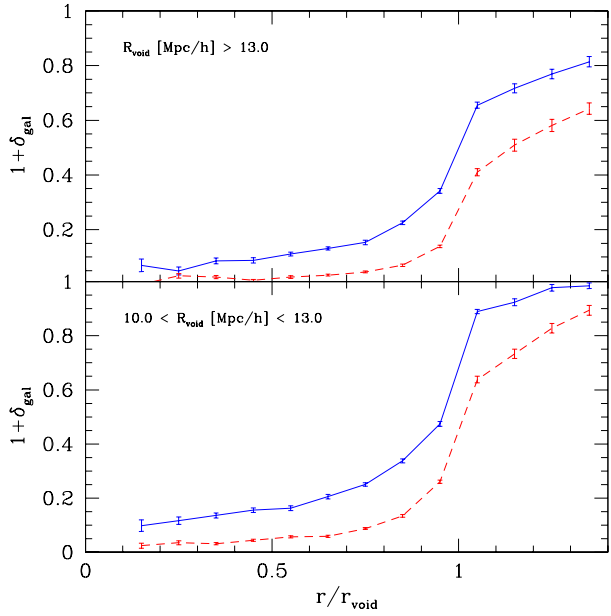}
\caption{Top panel: The local number density profile for red galaxies (dashed line) and
blue galaxies (full line) as a function of the normalized void radius 
for voids with radius larger than $13 \mpch$ found in the Millennium Run galaxy 
catalog. Bottom panel: the same profiles but for voids with radius between 
$10 \mpch$ and $13 \mpch$.}
\label{fig:fig7}
\end{figure}

In Figure \ref{fig:fig7c} we show the number density profile for galaxies in voids 
with radius larger than $13 \mpch$ (top panel) and
with radius between $10 \mpch$ and $13 \mpch$ (bottom panel) as a function of the normalized 
void radius (dashed line). 
As a comparison, we also show the same profiles found in the SDSS sample (full 
lines). 
In Figure \ref{fig:fig7} we plot the density profiles
separately for both blue and red galaxies, as we did for
the observed sample. The profiles from the simulations are similar to
those of the observed galaxies.

\begin{figure}
\includegraphics[width=\columnwidth]{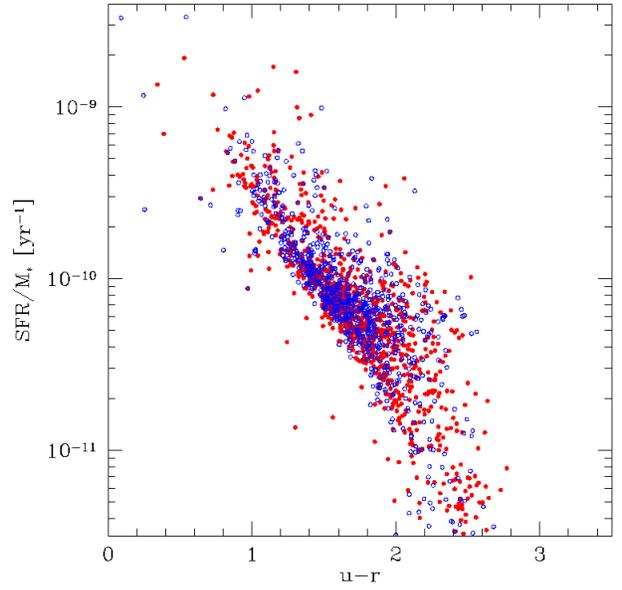}
\caption{Specific Star Formation Rate as a function of $u-r$ color for the simulated galaxies. 
Open circles denote galaxies in voids while filled circles are galaxies in the field.}
\label{fig:fig8}
\end{figure}

In Figure \ref{fig:fig8} we present the specific
SFR as a function of $u-r$ color for galaxies within the large voids in the Millennium
Run. This agrees well with that obtained for galaxies in our 
observed sample, with comparable star formation rates at fixed
color for void and field galaxies.

\begin{figure}
\includegraphics[width=\columnwidth]{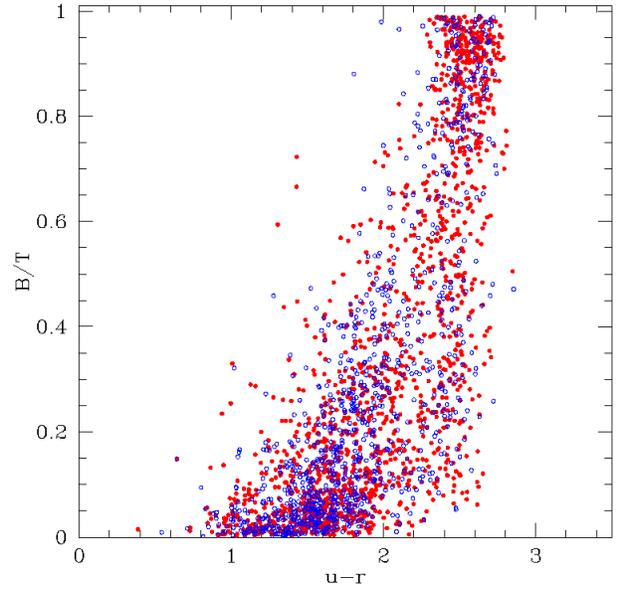}
\caption{$B/T$ ratios for the galaxies in voids as predicted by the
Millennium Run (open circles). Note that, blue galaxies are disk
dominated while red galaxies are bulge dominated. Red galaxies with
$u-r >2.5$ have $B/T$ ratios which suggest they are elliptical
galaxies. For a comparison we show the same $B/T$ ratios as a function
of color for galaxies in the field (filled circles). Galaxies in voids
 are similar to that of the field.}
\label{fig:fig9}
\end{figure}

\begin{figure}
\includegraphics[width=\columnwidth]{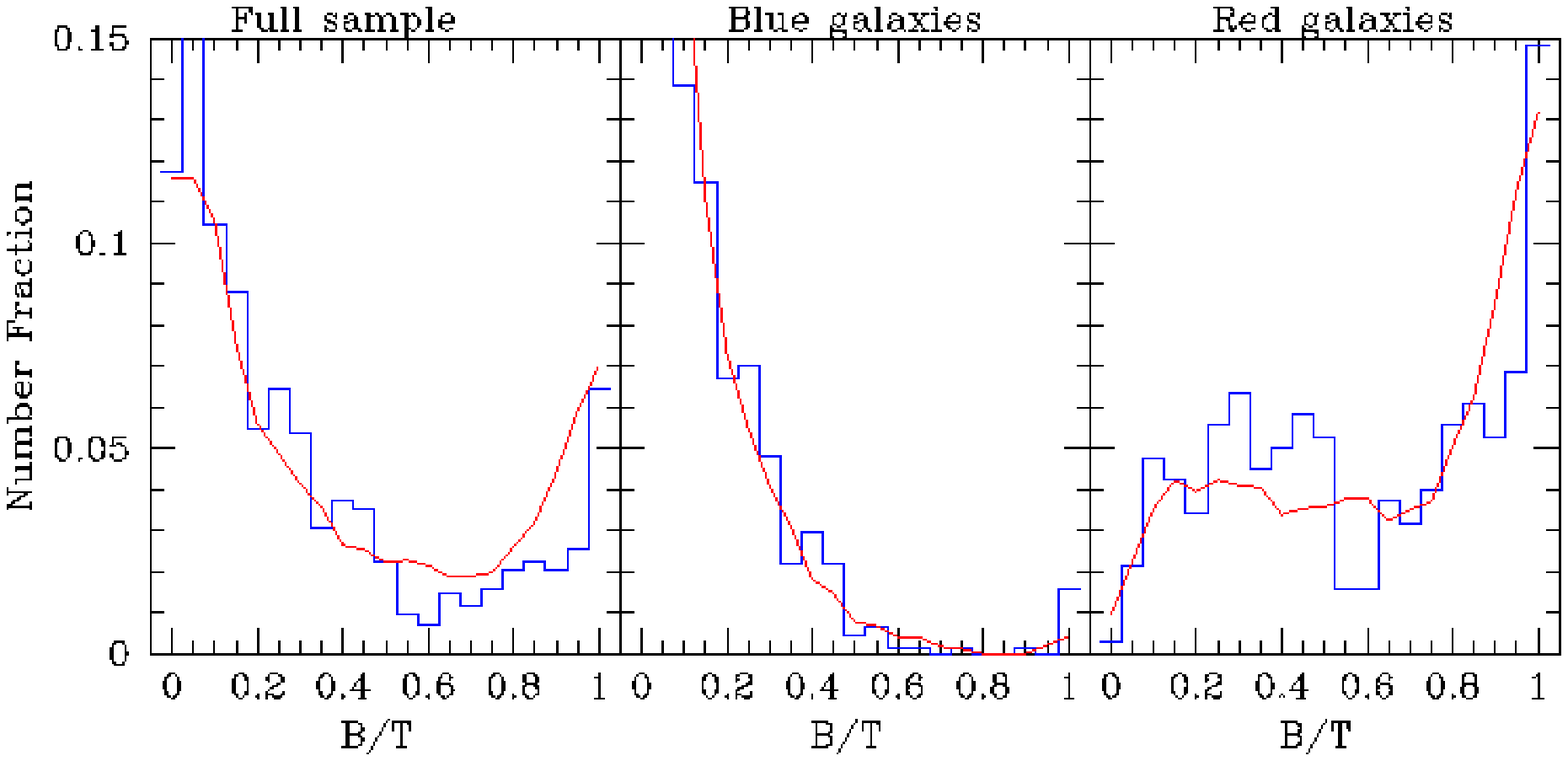}
\caption{$B/T$ distributions for the galaxies in voids (histograms) 
and field (curves) in the Millennium Run. Left panel: Full samples. 
Central Panel: Blue galaxies. Right panel: Red galaxies}
\label{fig:fig9b}
\end{figure}

We use the bulge and total magnitudes in the Millennium Run galaxy
catalog to compute the Bulge-to-Total luminosity ratio as a function
of the $u-r$ color for the galaxies inside the large voids (see Figure
\ref{fig:fig9}). One can see that blue galaxies are disk-dominated while
red galaxies are bulge-dominated. The B/T ratios of very red galaxies
(those with $u-r > 2.5$) suggest that these galaxies are ellipticals. 
The relation between morphology and color appears to be the same in
void galaxies as in field galaxies, although the relative number of
galaxies as a function of morphology or color differs. To see this 
more patent, we show in Figure \ref{fig:fig9b} the histograms of $B/T$
ratios for both void and field galaxies for the populations all together (left panel),
for the blue population (central panel) and red population (right panel)

\section{Discussion and Conclusions}

We use the SDSS Data Release 4 to study the properties of galaxies
lying in the most underdense regions of the universe. We search for
nearby ($z \leq 0.085$) large voids with radius larger than $10.0
\mpch$ in the SDSS North Strip. In total, we find 46 voids larger than
$10.0 \mpch$ with average underdensities $\delta_{gal} \sim -0.87$. We
study the $u-r$ color distributions, the qualitative morphology, and
the specific star formation rate (SFR/$M_{*}$) for a total of 495
galaxies with $-19.4+5\log h >M_{r}>-20.4 +5\log h$ lying in these
voids, and compare these properties with a control sample taken
randomly from the SDSS volume. It is important to realize that we do
not deal with luminous galaxies here, since by definition, our voids
have no luminous galaxies.

Galaxies in voids are not distributed randomly throughout the void. Typically, they
are found in filaments. In addition, the density increases toward the
void walls.  As a result, although the average galaxy density is very low 
in the voids, it is not necessarily as low {\it locally} compared with
galaxies outside voids.  The number density profiles for the largest voids
(those larger than $13 \mpch$) are flat in the inner part of voids and
increase rapidly as we go to outer parts, reaching about 60\% of the field
density at 1.4 void radius. The blue and red galaxies have similar profiles.
For smaller voids with radii between $10$ and $13 \mpch$, the
profiles are flat in the inner half of the voids and rise to the mean density 
at 1.2 void radii.  Recent results on the radial profiles of voids
defined by dark matter halos (Colberg et al. 2005; Patiri et al. 2006a)
show similar trends in the inner parts of voids, but suggest that
the density should increase suddenly just beyond 1 $R_{void}$
to form a prominent spike, after which the density declines and reaches the
mean density at about $1.6R_{void}$. This effect might be due to
crossing of spherical shells in the edges of voids. However, we do not see this
spike in our observations.

The $u-r$ color distribution in our luminosity range is clearly bimodal
in both the void and control samples. The fraction of blue galaxies,
however, is significantly larger in the voids, leading to a galaxy population
that is overall bluer. However, the effect does not appear to be as
 large as for samples that select  environment on a smaller scale
(e.g., Balogh et al 2004). For example, in the same magnitude range, 
Balogh et al. (2004) find significantly more blue than red galaxies in
their lower local density bins, while we find comparable numbers of
blue and red galaxies in our large voids.  It is impossible for us to
define local density as in Balogh et al. (2004), because their density
estimator is based on the proximity of galaxies with $M_r<-20$, of which
there are none in our voids, by definition. But the results suggest that,
if anything, there may be an excess of redder galaxies in voids over
what one might naively expect. 

If one considers the blue and red galaxies separately, there
does not appear to be any difference between void galaxies and the
overall population of galaxies. Within the blue and red populations separately, the
color distributions, morphology distributions (i.e. concentration and $B/T$ ratios), 
and correlation of specific star formation rate with color appear nearly 
indistinguishable between void and non-void galaxies. A similar result has been found
when galaxies are separated by their \textit{local} environment by
Balogh et al. (2004). These results suggest that the processes governing
the star formation history, which determines the color and specific
star formation rate, are largely independent of environment on any scale.

We compare our results with the latest predictions from
state-of-the-art semi-analytic models using the Millennium Run galaxy
catalog (Springel et al. 2005; Croton et al. 2005).  The predictions
from the Millennium Run qualitatively agree with the observed results.
The number density profiles for Millennium are in very good agreement
with those of the observations, showing the same trends. We do not see
a density spike at the edges of voids when we look at the profiles either in
real or redshift space.

In the Millennium run sample, as in the observed data, we see an
excess of blue galaxies and a deficit of red galaxies in voids with
respect to the overall galaxy population, but the effect is more
extreme in the simulations than in the observed data.  Within the blue
population, the simulated galaxies appear identical inside and outside
of voids, as we observe, but the red population appears to differ
between void and non-void galaxies, in contrast to what we observe.
For the specific star formation rates, we do not see any systematic
differences between galaxies in voids and the general
population. Still, the specific star formation rates seem more
concentrated in galaxies with u-r color around 1.6. Analysis of the
B/T relation for galaxies in the Millennium Run, again, does not show
any systematic difference between galaxies in voids and the
field. There are ellipticals in voids, still less than in the field.

\section*{Acknowlegments}

We would like to thank Darren Croton, Charlie Conroy and Simon D.M. White for useful 
comments and discussions that helped to improve the manuscript. 
SGP, FP \& JBR thank support from the Spanish MEC under grant PNAYA 2005-07789. AK
and JH acknowledge support of the NSF grant AST-0407072. SGP and FP
thank the Astronomy Department of NMSU for hospitality during their
visits to NMSU.

The Millennium Run simulation used in this paper was carried out by the
Virgo Supercomputing Consortium at the Computing Center of the Max-Planck
Society in Garching.

Funding for the creation and distribution of the SDSS Archive has
been provided by the Alfred P. Sloan Foundation, the Participating
Institutions, the National Aeronautics and Space Administration,
the National Science Foundation, the U.S.  Department of Energy, the
Japanese Monbukagakusho, and the Max Planck Society.  The SDSS Web site
is http://www.sdss.org/.

The SDSS is managed by the Astrophysical Research Consortium (ARC)
for the Participating Institutions. The Participating Institutions are
The University of Chicago, Fermilab, the Institute for Advanced Study,
the Japan Participation Group, The Johns Hopkins University, the Korean
Scientist Group, Los Alamos National Laboratory, the Max-Planck-Institute
for Astronomy (MPIA), the Max-Planck-Institute for Astrophysics (MPA),
New Mexico State University, University of Pittsburgh, University of
Portsmouth, Princeton University, the United States Naval Observatory,
and the University of Washington.



\begin{thebibliography}{}

\bibitem{} Abazajian, K., et al. 2003, \aj, 126. 2081
\bibitem{} Baldry, I. K., et al. 2004, \apj, 600, 681
\bibitem{} Benson, A. J., Hoyle, F., Torres, F., \& Vogeley, M. S. 
 2003, MNRAS, 340, 160 
\bibitem{} Balogh, M. L., et al. 2004, \apj, 615, L101
\bibitem{} Bernardi, M., et al. 2003a, \aj, 125, 1849
\bibitem{} Blanton, M. R., et al. 2001, \apj, 121, 2358
\bibitem{} Blanton, M. R., Lin, H., Lupton, R. H., Maley, F. M., Young, N.,
 Zehavi, I., \& Loveday, J. 2003a, \aj, 125, 2276
\bibitem{} Brinchmann, J., et al. 2004, \mnras, 351, 1151
\bibitem{} Colberg, J.~M., Sheth, R.~K., Diaferio, A., Gao, L.,
 \& Yoshida, N.\ 2005, \mnras, 360, 216 
\bibitem{} Croton, D., et al. 2005, \mnras, 356, 1155
\bibitem{} Dressler, A. 1980, \apj, 236, 351
\bibitem{} Dressler, A, et al. 2004, \apj, 617, 867 
\bibitem{} Eisenstein, D. J., et al. 2001, \aj, 122, 2267
\bibitem{} Fukugita, M., Ichikawa, T., Gunn, J. E., Doi, M., 
 Shimasaku, K., \& Schneider, D. P. 1996, \aj, 111, 1748
\bibitem{} Grogin, N. A., \& Geller, M. J. 2000, \apj, 119, 32
\bibitem{} Gottl\"ober, S. et al. 2003, \mnras, 344, 715
\bibitem{} Gunn, J. E., et al. 1998, \apj, 116, 3040
\bibitem{} Hogg, D. W., et al. 2004, \apj, 601, L29
\bibitem{} Kauffmann, G., et al. 2003, \mnras, 341, 33
\bibitem{} Lupton, R. H., Gunn, J. E., \& Szalay, A. S. 1999, \aj, 118, 1406
\bibitem{} MacArthur, L.~A., Courteau, S., \& Holtzman, J.~A. 2003, \apj, 582, 689 
\bibitem{} Patiri, S.~G., Betancort-Rijo, J., \& Prada, F. \ 2006a, \mnras, 368, 1132
\bibitem{} Patiri, S. G., Betancort-Rijo, J. E., Prada, F., Klypin, A. K. 
 \& Gottl\"ober, S. \ 2006b, \mnras, 369, 335
\bibitem{} Pier, J. R., et al. 2003, \aj, 125, 1559
\bibitem{} Postman, M., \& Geller, M. J. 1984, \apj, 281, 95
\bibitem{} Rojas, R.~R., Vogeley, 
  M.~S., Hoyle, F. \& Brinkmann, J.\ 2004, \apj, 617, 50 
\bibitem{} Rojas, R.~R., Vogeley, 
  M.~S., Hoyle, F. \& Brinkmann, J.\ 2005, \apj, 624, 571 
\bibitem{} Scranton, R. et al. 2002, \apjs, 579, 48 
\bibitem{} Sheth, R. \& Tormen, G., 2002, \mnras, 39, 61
\bibitem{} Springel, V. et al. 2005, {\emph Nature}, 435, 629
\bibitem{} Stoughton, C. et al. 2002, \aj, 123, 485
\bibitem{} Strauss, M. A., et al. 2002, \aj, 124, 1810S
\bibitem{} Strateva, I., et al. 2001, \apj, 122, 1874
\bibitem{} York, D. G. et al. 2000, \aj, 120, 1579




\end{thebibliography}
\end{document}